\documentclass[3p]{elsarticle}

\usepackage{lineno,hyperref}
\modulolinenumbers[1]
\usepackage{graphicx}
\usepackage{subfigure}

\usepackage[nameinlink]{cleveref}
\usepackage{array}
\usepackage{multirow}
\usepackage{tabularx}
\usepackage{ulem}
\usepackage{xcolor}

\journal{Journal of \LaTeX\ Templates}

\begin{document}

\begin{frontmatter}

\title{Effect of color reconnection on multiplicity dependent charged particle production in PYTHIA-generated pp collisions at the LHC energies}

\author[1,2]{Pranjal Sarma}

\ead{pranjalsarmaedu@gmail.com}
\address[1]{Nuclear and Radiation Physics Research Laboratory, Department of Physics, Gauhati University, \\Guwahati, 781014, Assam,  India}
\address[2]{Department of Physics, Jagiroad College, Jagiroad, 782410, Assam, India}

\author[1]{Banajit Barman}
\author[1]{Buddhadeb Bhattacharjee \corref{mycorrespondingauthor}}


\cortext[mycorrespondingauthor]{Corresponding author}
\ead{buddhadeb.bhattacharjee@cern.ch}


\begin{abstract}

The transverse momentum ($p_{\rm T}$) spectra of inclusive charged particles and its dependence on charged-particle multiplicity are studied in pp collisions at $\sqrt{s}$ = 7 and 13 TeV using the Monte Carlo event generator PYTHIA Monash. The results of the minimum bias pp collisions are compared with the ALICE published results at the same energies. The variations of the effective temperature (T$_{\rm {Eff}}$) and average transverse momentum ($\langle p_{\mathrm{T}} \rangle$) with charged-particle multiplicity are also studied. The $p_{\rm T}$ spectra of the charged particles are observed to get harder with increasing charged-particle multiplicity. Moreover, a sharp increase in the T$_{\rm {Eff}}$ and $\langle p_{\mathrm{T}} \rangle$ with increasing multiplicity followed by a gradual decrease with the further increase in multiplicity could also be observed. Further, it could be observed that the color reconnection (CR) mechanism has a considerable effect on flattening the T$_{\rm {Eff}}$ and $\langle p_{\mathrm{T}} \rangle$ with increasing multiplicity, which indicates that phase-transition may not necessarily be the only mechanism that could give rise to such an effect.

\end{abstract}


\end{frontmatter}


\section{Introduction}

Relativistic heavy-ion collisions provide a unique opportunity to investigate the properties of matter created under extreme conditions of temperature and density, in which, the degrees of freedom of the matter under exploration might change from hadronic to partonic resulting Quark-Gluon Plasma (QGP) \cite{pbm, jharris}. Different experiments, namely SPS at the CERN \cite{heinz} and RHIC at the BNL \cite{arsene, adcox, bback, jadams_star, shuryak, huovinen, muller}, have been designed to create and characterize QGP in the laboratory. Several results from these experiments have already provided significant indications of the formation of the QGP state, and further, its existence has also been confirmed by the latest results from different experiments of LHC at the CERN \cite{abelev1, jadam_alice1, jadam_alice2, abelev2, PhysRevC.84.024906}. Contrary to this, the proton-proton (pp) system, with a few valence partons for the collisions, has traditionally been used as reference measurements for the heavy-ion collisions. However, at the LHC energies, the high-multiplicity pp events can be compared with those of the $p$-Pb and peripheral Pb-Pb collisions. And there are several evidences that the matter formed in such events exhibits collective-like effects providing hints that such collective behaviors are not the characteristic of heavy-ion collisions only \cite{bzdak, jadam_alice3, antonio, pranjal}. As the pp collisions at the LHC energies are believed to be collisions of two composite structures consisting of large numbers of partons \cite{proton_lhc1, proton_lhc2}, several workers opined that the collisions of such composite structure might give rise to a thermally equilibrated system \cite{jadam_alice3, qgp_small}.

The PYTHIA Monte Carlo (MC) event generator is a general-purpose pQCD based event generator that is capable of describing different experimental results of pp collisions quite successfully at the LHC energies \cite{jadam_alice4, acharya_alice1, abelev4}. It uses factorized perturbative expansion for the hard parton-parton scattering, initial and final state partons showers, various models for hadronization, and multiparton interactions \cite{pythia_6}. Different parameters of the PYTHIA model have been improved or tuned from time to time by comparing the available experimental results for a better description of data as well as for a more in-depth understanding of the underlying collision dynamics \cite{skands}.

The transverse momentum ($p_{\mathrm{T}}$) spectra of the charged particles produced in collisions carry information about the collisions dynamics and full space-time evolution of the system from initial to the final stage of the collisions \cite{adler}. The flattening of the $p_{\mathrm{T}}$ spectra with multiplicity is traditionally considered as an indication of the onset of formation of a mixed phase of de-confined partons and hadrons. In the hydrodynamical model, the inverse slope of the $p_{\mathrm{T}}$ spectra is considered as a measure of the combined effect of the kinetic freeze-out temperature and transverse expansion flow of the system \cite{vhove}. Thus, studies on the $p_{\mathrm{T}}$ spectra provide information about the effective temperature (T$_{\rm {Eff}}$) of the system. Observation of a plateau-like region in the variation of T$_{\rm {Eff}}$ against multiplicity is considered as a possible signal of formation of a mixed-phase, similar to that observed in the change of temperature with entropy for the first-order phase-transition. Moreover, studies on the average transverse momentum ($\langle p_{\mathrm{T}} \rangle$) with multiplicity are often carried out to realise a possible phase-transition from QGP to hadron. Here again, the flattening of $\langle p_{\mathrm{T}} \rangle$ with multiplicity is considered to be an indication of QGP like phase-transition \cite{vhove}. In this work, an attempt has therefore been made to study the dependence of the $p_{\mathrm{T}}$ spectra, T$_{\rm {Eff}}$ and $\langle p_{\mathrm{T}} \rangle$ of the charged particles for different multiplicity interval in pp collisions at $\sqrt{s}=$ 7 and 13 TeV using the MC event generator PYTHIA Monash \cite{skands}. 

PYTHIA also includes color reconnection (CR) mechanism, a string fragmentation model that considers final partons to connect through their color charge, in such a way that the total string length becomes minimum. Therefore, two partons produced from independent hard scattering are color connected and make a large transverse boost \cite{gustafson}. Several recent experimental results of pp collisions, particularly of high-multiplicity events, have shown a number of collective behaviors such as enhanced production of baryon over meson at intermediate $p_{\rm T}$, mass-dependent growth in $\langle p_{\rm T} \rangle$ with multiplicity, etc., similar to those observed in heavy-ion collisions \cite{abelev1, antonio}. However, when compared with the PYTHIA generated MC data, the CR mechanism was found to mimic a number of such collective-like effects, although the results of the PYTHIA model with CR are not in quantitative agreement with the experimental observations \cite{jadam_alice3, pp7tev, acharya_pikpvsmult}. Further, more recently, it has been observed that the CR mechanism in PYTHIA has significant effects on the intermittent behavior of charged particles produced in high-multiplicity pp collisions \cite{pranjal}. It, therefore, becomes very much pertinent to investigate the role of the color reconnection on other global signatures of collective behaviors of heavy-ion collisions. Inspired by this argument, in this work, a detailed multiplicity dependent study on global observables of hadron-hadron collisions, in particular, T$_{\rm {Eff}}$ $\&$ $\langle p_{\mathrm{T}} \rangle$, for different strengths of the CR mechanism has been undertaken to access the contribution of color reconnection on the signature of collective behavior of nuclear collisions.

\section{Results and discussions}

For this investigation, 125 M $\&$ 73 M events were generated for pp collisions at $\sqrt{s}=$~7 and 13 TeV, respectively, using the PYTHIA Monash (default) event generator and are analyzed.

\begin{figure*}[htp]
\centering
	\subfigure[]{\includegraphics[width=57mm, height=67mm]{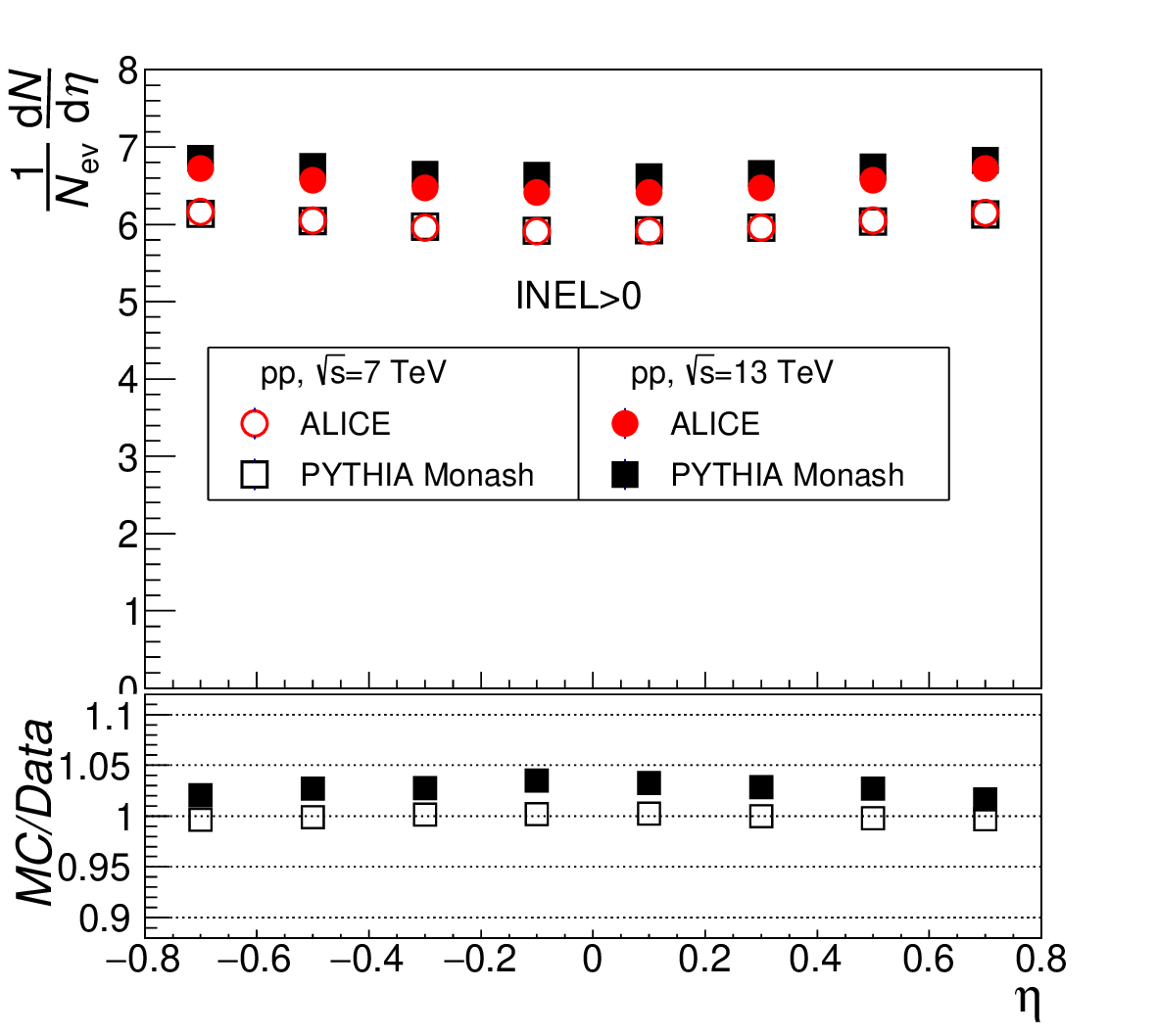}\label{fig:etadist}}
	\subfigure[]{\includegraphics[width=57mm, height=67mm]{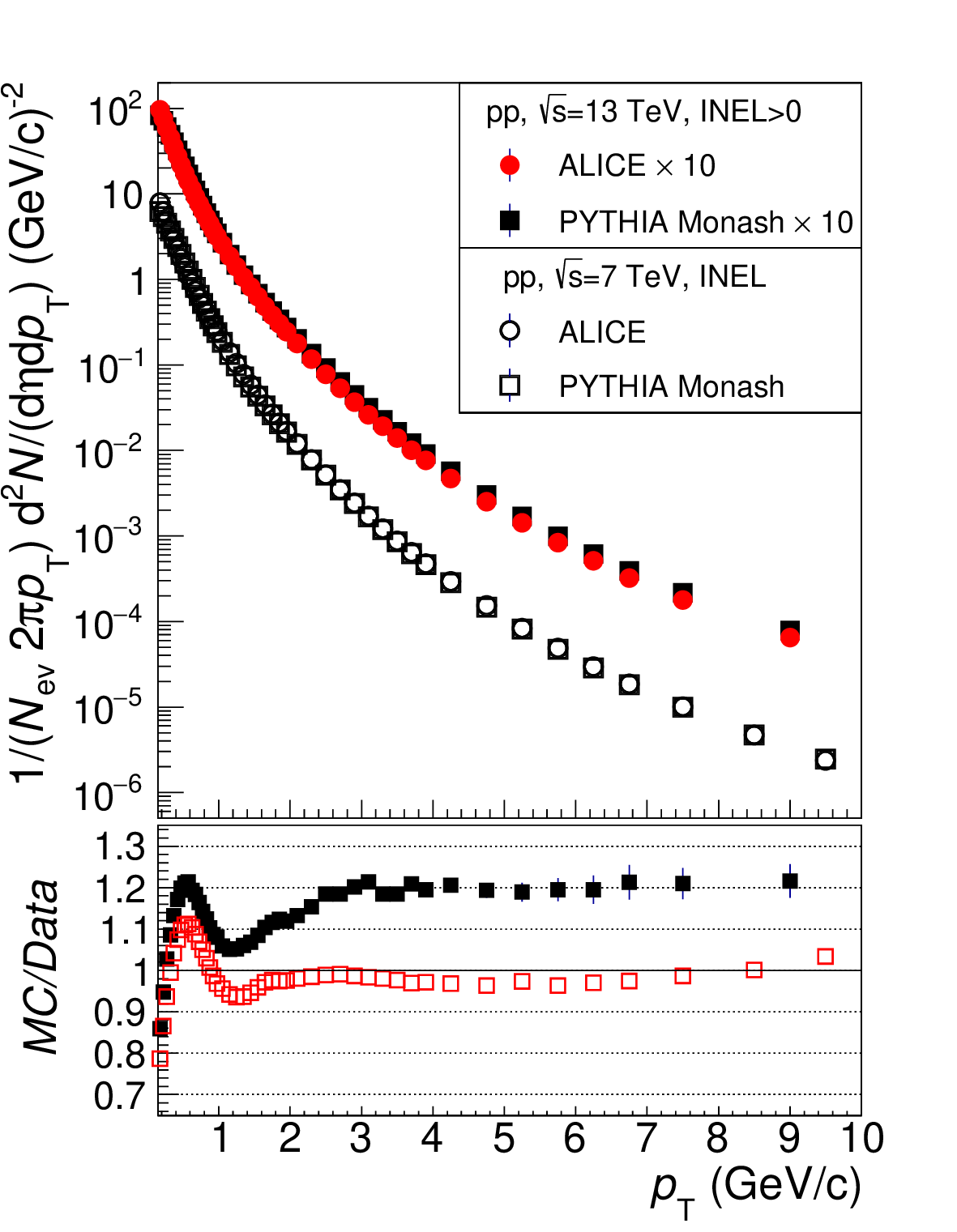}\label{fig:ptdist}}
	\caption{(a) Pseudorapidity and (b) $p_{\mathrm{T}}$ distributions of the charged particles in PYTHIA Monash generated pp collisions at $\sqrt{s}=$ 7 and 13 TeV are compared with the published results of the ALICE Collaboration \cite{jadam_alice4, jadam_alice5, abelev3}.}
	\end{figure*}

As a quality assurance test of our generated sets of data, within the acceptance of the ALICE detector, the pseudorapidity distributions of the primary charged particles in minimum bias (MB) pp collisions at $\sqrt{s}=$ 7 and 13 TeV are plotted in Fig. \ref{fig:etadist} and compared with the published results of the ALICE Collaboration \cite{jadam_alice4, jadam_alice5}. From the figure, it is readily evident that the pseudorapidity distributions of the PYTHIA Monash generated data are in good agreement with the ALICE experimental results.

In Fig. \ref{fig:ptdist}, the transverse momentum distributions of the primary charged particles with PYTHIA Monash generated data are plotted for minimum bias pp collisions at $\sqrt{s}=$ 7 and 13 TeV and compared with the published result of ALICE \cite{ jadam_alice5, abelev3}. It could be observed from the figure that for pp collisions at $\sqrt{s}=$ 7 TeV, except for the low $p_{\mathrm{T}}$ region ($<$0.8 GeV/c), the PYTHIA Monash and ALICE results are in very good agreement in the studied $p_{\mathrm{T}}$ range. Further, the disagreement between the PYTHIA generated and experimental data of ALICE for $\sqrt{s}=$ 13 TeV is also found to be not very large and remains within the limit of 20$\%$. It is important to note that the ALICE published $p_{\mathrm{T}}$ spectra for pp collisions at $\sqrt{s}=$ 7 TeV is measured for inelastic events (INEL) \cite{abelev3}, whereas, for $\sqrt{s}=$ 13 TeV, the spectra is measured for events having at least one charged particle produced with $p_{\mathrm{T}}$$>$0 in $|\eta|$$<$1 (INEL$>$0) \cite{jadam_alice5}. As the tuning of different parameters of the PYTHIA Monash has performed for inelastic events, better agreement of the $p_{\mathrm{T}}$ spectra in pp collisions at $\sqrt{s}=$ 7 TeV is justifiable \cite{skands}.

The charged particle multiplicity (N$_{\rm {ch}}$) provides central information about the particle production mechanism and is considered as an important global observable of nuclear collisions \cite{beggio, dremin}. As the PYTHIA Monash generated sets of data are quite successful in describing the results of pseudorapidity and $p_{\mathrm{T}}$ spectra of ALICE at the studied energies; it is therefore expected that a study on the $p_{\mathrm{T}}$ spectra, T$_{\rm {Eff}}$ and $\langle p_\mathrm{T} \rangle$ of the primary charged particles for different charged-particle multiplicity would provide better insight about pp collision dynamics. 

\begin{figure}[htp]
	\centering
	\includegraphics[width=70mm, height=55mm]{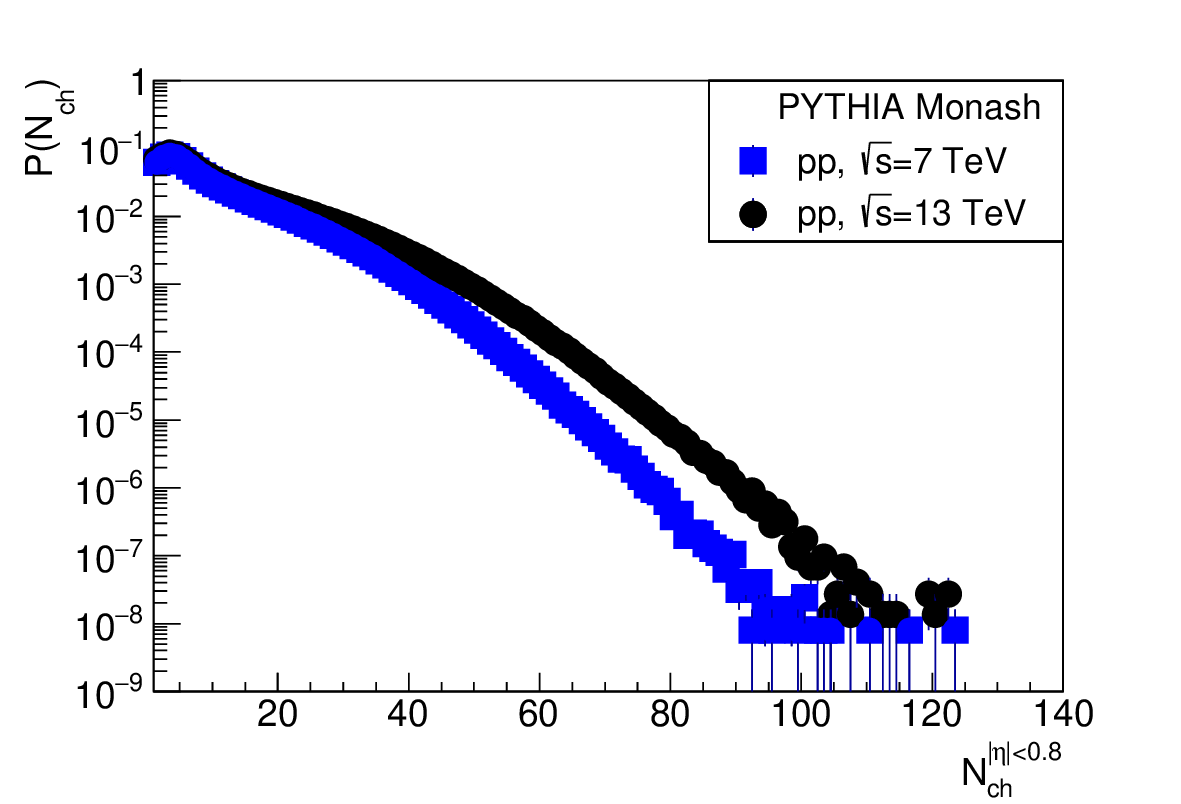}
	\caption{Multiplicity distributions of the primary charged particles produced within $|\eta|$$<$0.8 in pp collisions at $\sqrt{s}=$ 7 and 13 TeV for PYTHIA generated data.}
	\label{fig:multdist}
	\end{figure}

Multiplicity distribution of the charged particles produced in pp collisions at $\sqrt{s}=$7 and 13 TeV within the mid-pseudorapidity range ($|\eta|$$<$0.8) are shown in Fig. \ref{fig:multdist}. From the figure, it is seen that the charged-particles multiplicity increases with increasing center-of-mass energy. At LHC energies, pp collisions are considered to be collision of two balls of partons  \cite{proton_lhc1, proton_lhc2}. The high-multiplicity events occur at a small impact parameter, and therefore, the number of particles thus produced is expected to be higher due to the large energy-momentum transfer as well as for the increased number of multiparticle collisions \cite{barshay1, barshay2}. It is quite natural that the $p_{\mathrm{T}}$ spectra of the particles in such events also get affected, and a change in the behavior of the spectra is obvious, if studied as a function of charged-particle multiplicity.

\begin{figure}[htp]
	\centering
	\subfigure[]{\includegraphics[width=60mm, height=70mm]{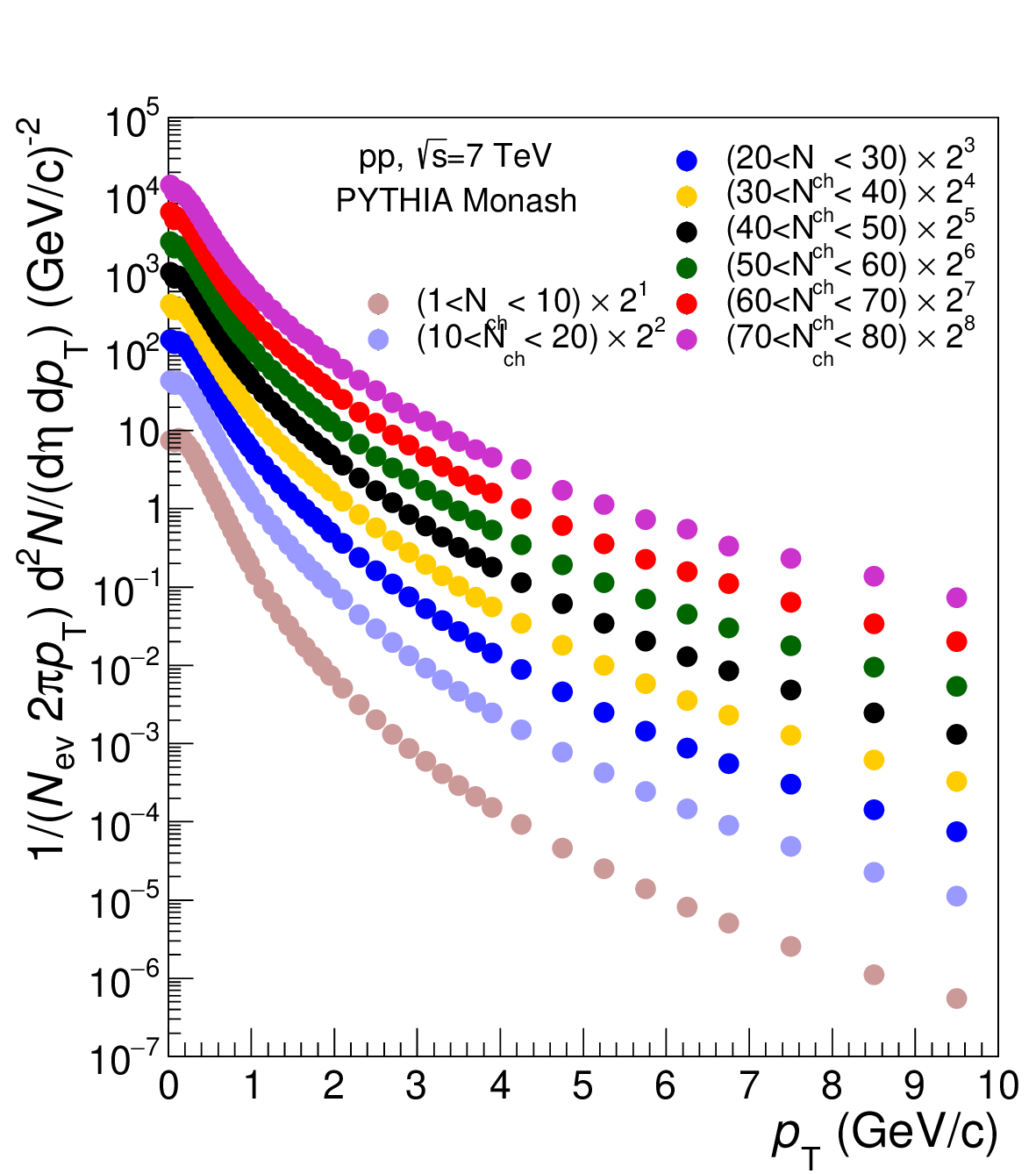}\label{fig:ptmult7}}
	\subfigure[]{\includegraphics[width=60mm, height=70mm]{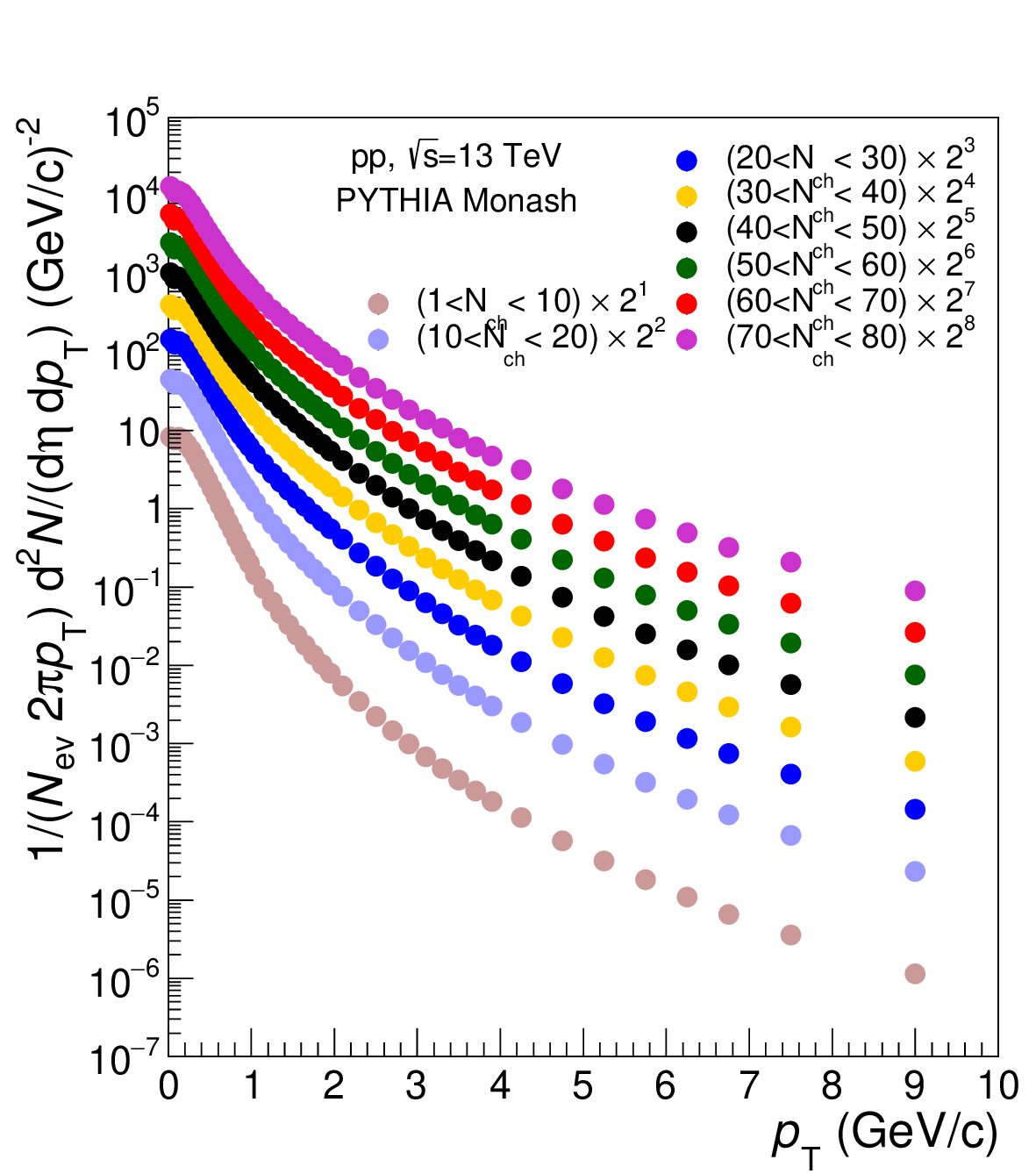}\label{fig:ptmult13}}
	\caption{$p_{\mathrm{T}}$ distributions of the PYTHIA generated charged particles as a function of charged-particle multiplicity in pp collisions at (a) $\sqrt{s}=$ 7 and (b) $\sqrt{s}=$ 13 TeV. $p_{\mathrm{T}}$ spectra are scaled by a factor of 2$^n$ (n=1, 2, ..., 8) for better visibility.}
	\end{figure}

In Figs. \ref{fig:ptmult7} $\&$ \ref{fig:ptmult13}, the $p_{\mathrm{T}}$ spectra of the charged particles for different charged-particle multiplicity intervals are plotted for pp collisions at $\sqrt{s}=$ 7 and 13 TeV, respectively with our generated sets of data. The charged particle $p_{\mathrm{T}}$ spectra are observed to become harder with increasing charged-particle multiplicity for both the studied energies. This characteristic behavior is similar to that proposed by Van Hove, in which, it is argued that the flattening of the $p_{\mathrm{T}}$ spectra with increasing multiplicity might indicate the de-confinement transition of hadronic matter \cite{vhove}.

At low $p_{\rm T}$, the charged particles transverse momentum spectra can be parametrized by Eq. \ref{eq:eq1} \cite{Gorenstein:2003cu}. 

\begin{equation}
\label{eq:eq1}
\frac{\mathrm{d} N}{p_{\mathrm{T}}dp_{\mathrm{T}}} \sim C \times exp(-\frac{p_{\mathrm{T}}}{ \rm{T}_{\rm{Eff}}})
\end{equation}

Where T$_{\rm {Eff}}$ is the effective temperature of the system containing the contribution from both the kinetic freeze-out temperature and transverse expansion flow.

\begin{figure*}[htp]
	 \centering
	\subfigure[]{\includegraphics[width=65mm, height=54mm]{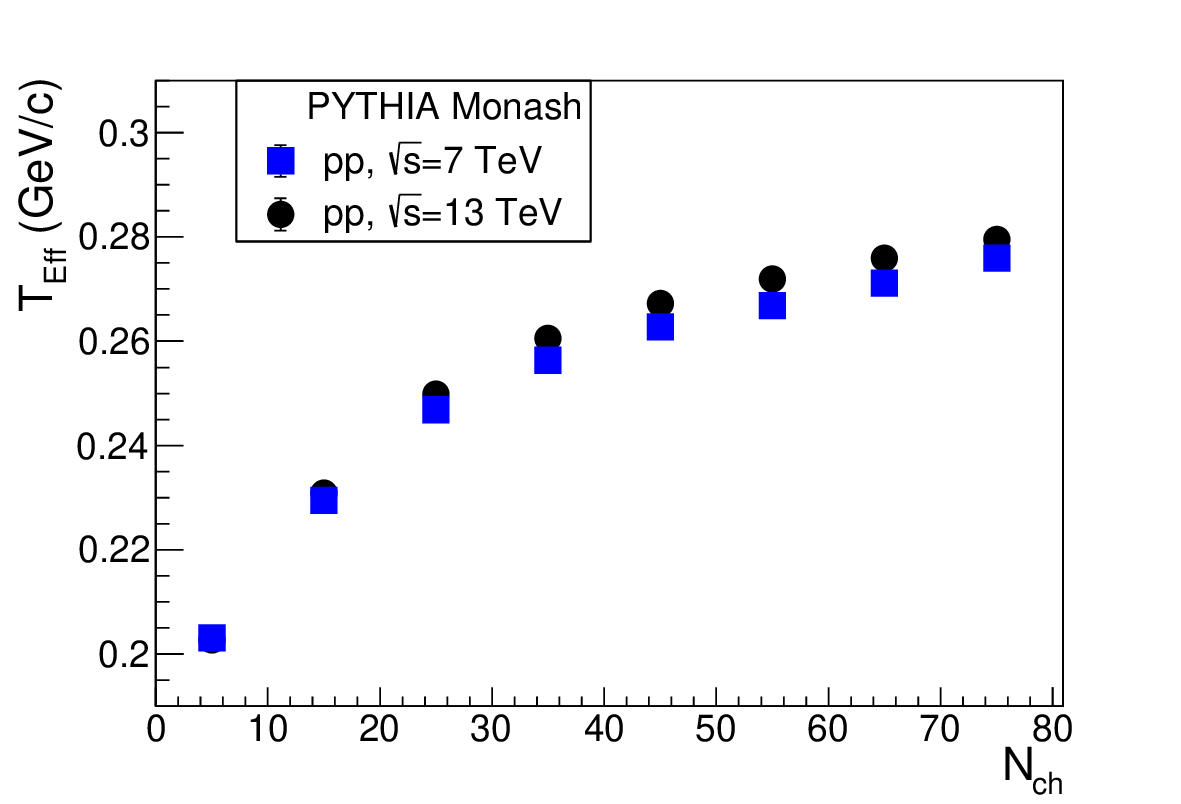}\label{fig:temp1}}
	\subfigure[]{\includegraphics[width=65mm, height=54mm]{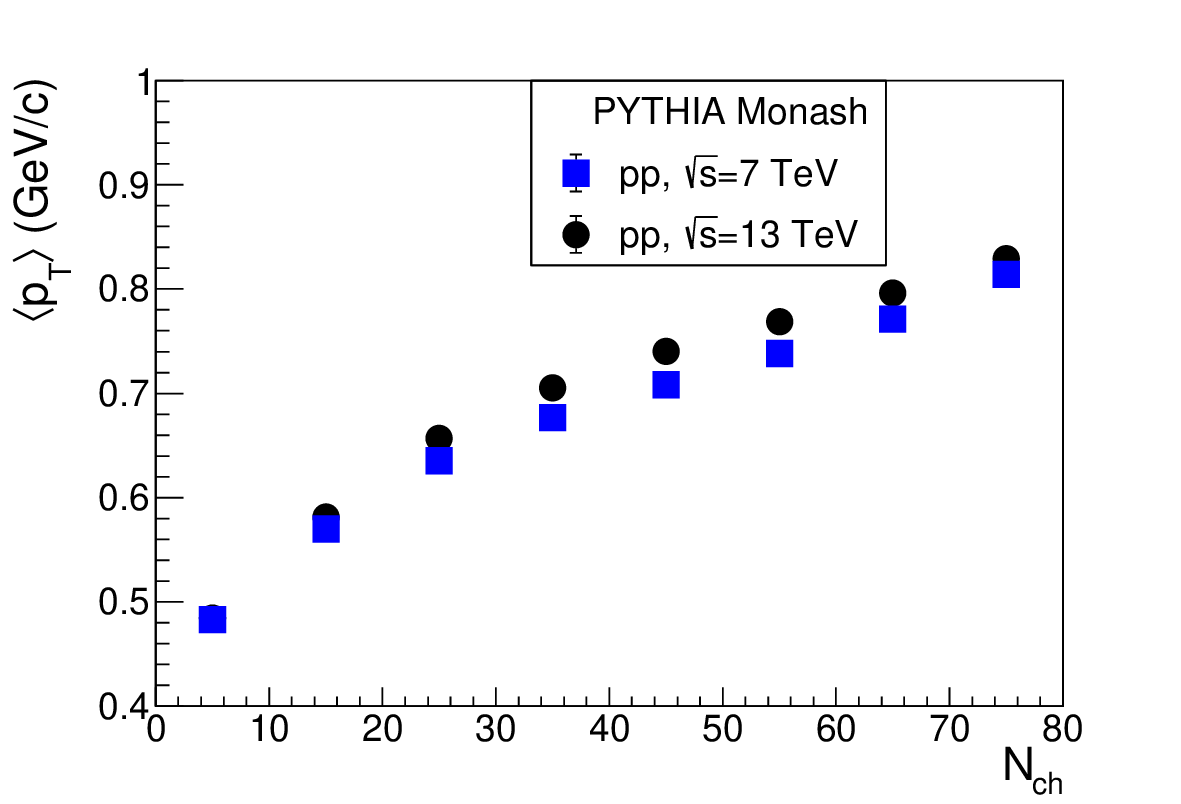}\label{fig:mpt1}}
	\caption{Variation of (a) effective temperature and (b) average $p_{\mathrm{T}}$ as a function of charged-particle multiplicity in pp collisions at $\sqrt{s}=$ 7 and 13 TeV for PYTHIA generated data.}
	\end{figure*}

In Figs. \ref{fig:temp1} $\&$ \ref{fig:mpt1}, the variations of the T$_{\rm {Eff}}$ and $\langle p_{\mathrm{T}} \rangle$ for different charged-particle multiplicity have been plotted respectively with the PYTHIA Monash generated data of pp collisions at $\sqrt{s}$ = 7 and 13 TeV. It is interesting to observe from these figures that both the T$_{\rm {Eff}}$ and $\langle p_{\mathrm{T}} \rangle$ increases sharply with the increase of charged-particle multiplicity up to $(\mathrm{N}_{\rm {ch}})$$<$30. However, with the further increase of multiplicity, the increasing behavior of the T$_{\rm {Eff}}$ and $\langle p_{\mathrm{T}} \rangle$ gradually slow down and tend to reach a plateau-like region, particularly in T$_{\rm {Eff}}$ vs. multiplicity plot. As PYTHIA does not include hadron to QGP phase-transition, the appearance of this plateau-like region for the PYTHIA-generated data suggests that apart from such transition, some other mechanisms, which may not necessarily be related to phase-transition could also give rise to such an effect. Further, T$_{\rm {Eff}}$ and $\langle p_{\mathrm{T}} \rangle$ do not show any significant dependence on the center-of-mass energy, which suggest that as long as the multiplicity remains the same, T$_{\rm {Eff}}$ and $\langle p_{\mathrm{T}} \rangle$ of the system do not change.

Color reconnection (CR) is found to give several collective-like behaviors, as observed in the heavy-ion collisions, in pp data \cite{abelev1, antonio}. Thus, studies on the effect of CR on transverse momentum and other related observables might shed light on collisions dynamics of the system under investigation. The parameter that determines the strength of the color reconnection in PYTHIA is known as the reconnection range (RR). To see the effect of the color reconnection mechanism on our studied observables, three new sets of data of 16.6~M, 56 M and 71 M events with RR=0.0, 3.0 and 3.6, respectively, are generated at the highest collisions energy available at the LHC for pp collisions, i.e., $\sqrt{s}=$ 13 TeV and further analysis is carried out with these datasets only.

\begin{figure}[htp]
	\centering
	\includegraphics[width=63mm, height=77mm]{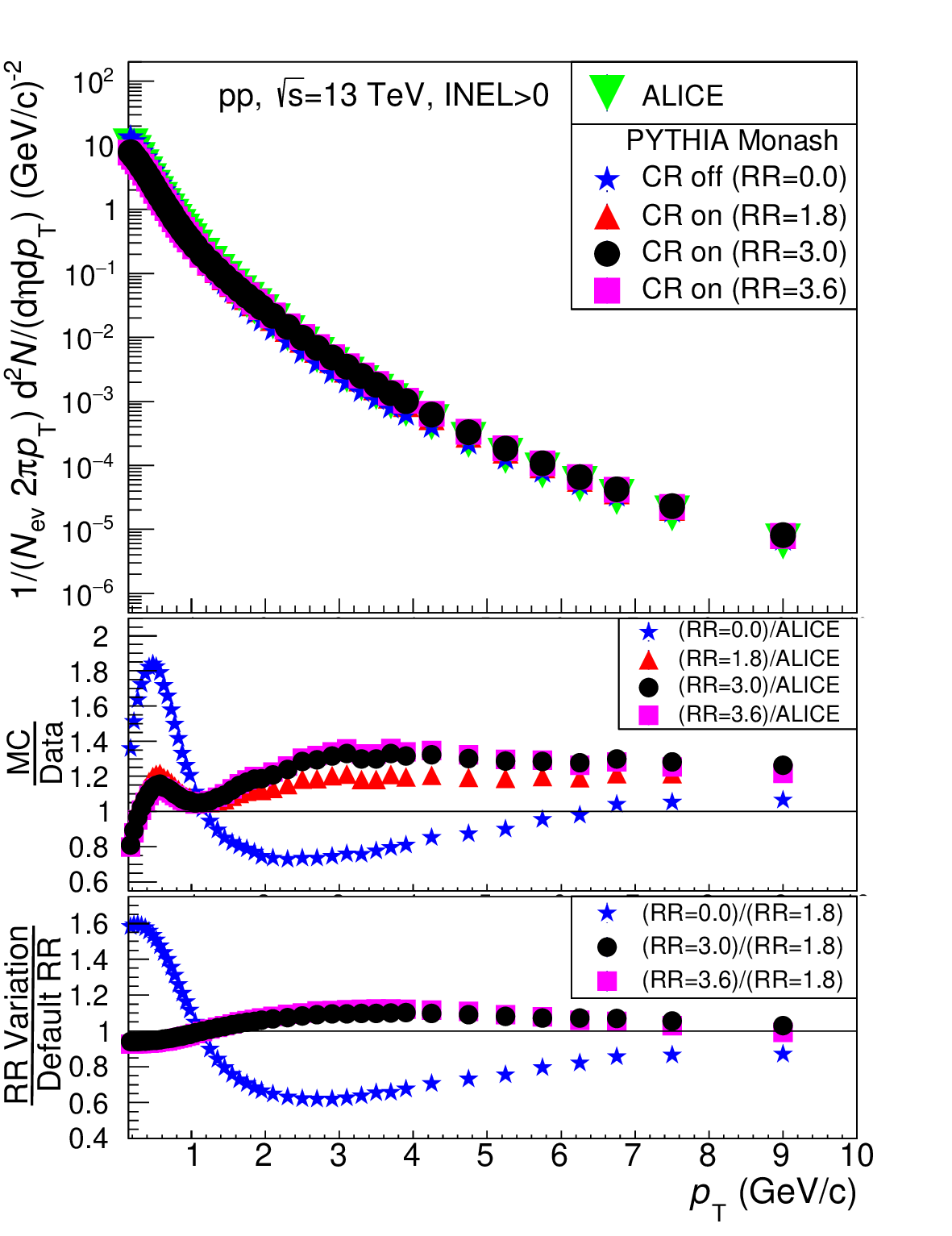}
	\caption{ (Top panel) $p_{\mathrm{T}}$ spectra of the charged particles of PYTHIA-generated data with different RR parameter are compared with the ALICE data for pp collisions at $\sqrt{s}=$ 13 TeV \cite{jadam_alice5}. (Middle panel) Ratio of $p_{\mathrm{T}}$ spectra of different RR parameter to ALICE data and (lower panel) ratio of $p_{\mathrm{T}}$ spectra of different RR to default RR value in PYTHIA. }
	\label{fig:rrcompspectra}
	\end{figure}

The $p_{\rm T}$ spectra obtained from the PYTHIA generated data in pp collisions at $\sqrt{s}=$~13~TeV for different values of the RR parameter are compared with the ALICE data, and the results are shown in the upper panel of Fig. \ref{fig:rrcompspectra}. From the ratio plots (middle panel), it is observed that in the low $p_{\rm T}$ region, CR off data overestimates (up to 85$\%$) the ALICE results, whereas, it underestimates (up to 29$\%$) in the intermediate $p_{\rm T}$ region. On the other hand, the RR=3.0 and 3.6 generated data provides a little better description of the ALICE data than the default RR data at low $p_{\rm T}$, although it become worse at the intermediate $p_{\rm T}$.

In the lower panel of Fig. \ref{fig:rrcompspectra}, a comparison of $p_{\rm T}$ spectra ratios for different values of the RR parameter to default RR is shown. From these ratio plots, it is readily evident that with CR off, particles tend to produce at low $p_{\rm T}$ compared to that of intermediate and high $p_{\rm T}$ regions. However, for higher reconnection range values (RR=3.0 $\&$ 3.6), particle production favors at the intermediate $p_{\rm T}$ than the low $p_{\rm T}$ region. This enhancement of charged particles for RR=3.0 and 3.6, is similar to those observed for the p/$\pi$ ratio in Ref. \cite{antonio} with the default CR (RR=1.5) in PYTHIA 4C tune. This usually occurs because of the experience of a higher boost by the particles due to increase in the strength of the CR mechanism, resulting in the shift of particles from a low to a comparatively higher $p_{\rm T}$ region.

\begin{figure}[htp]
	\centering
	\subfigure[]{\includegraphics[width=65mm, height=53mm]{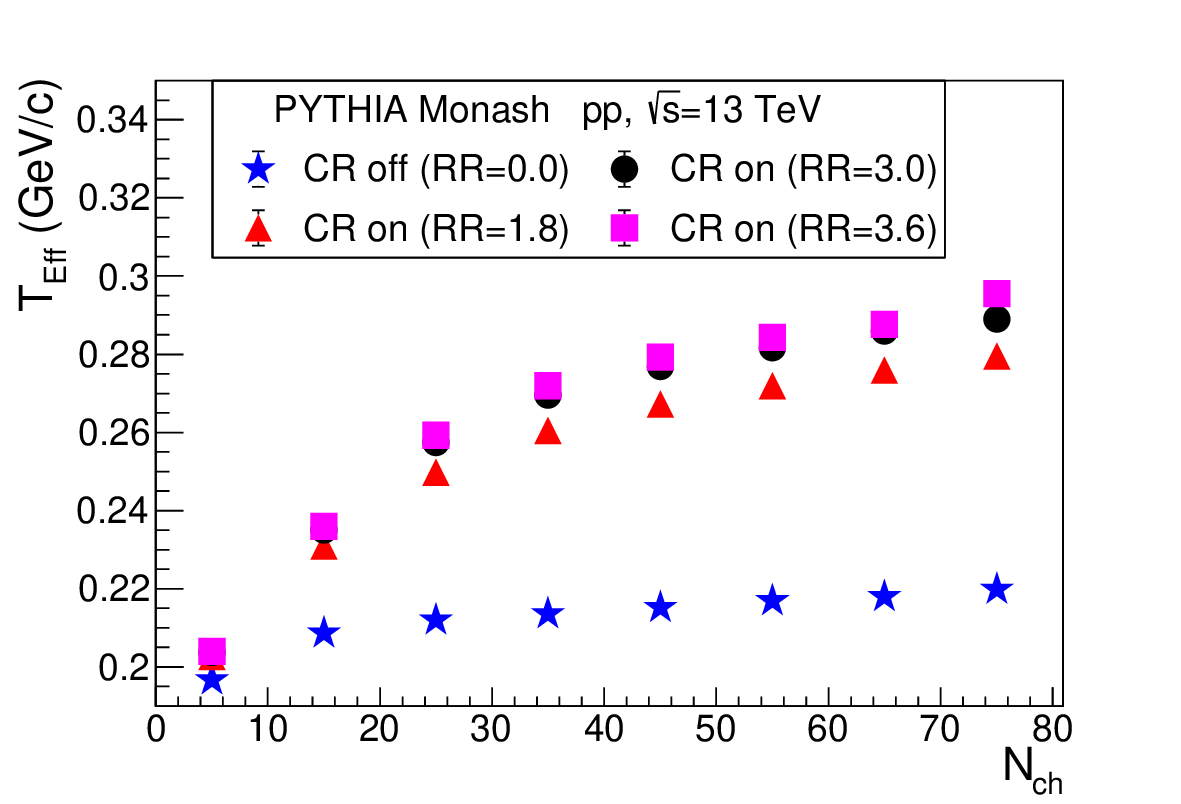}\label{fig:temp2}}
	\subfigure[]{\includegraphics[width=65mm, height=53mm]{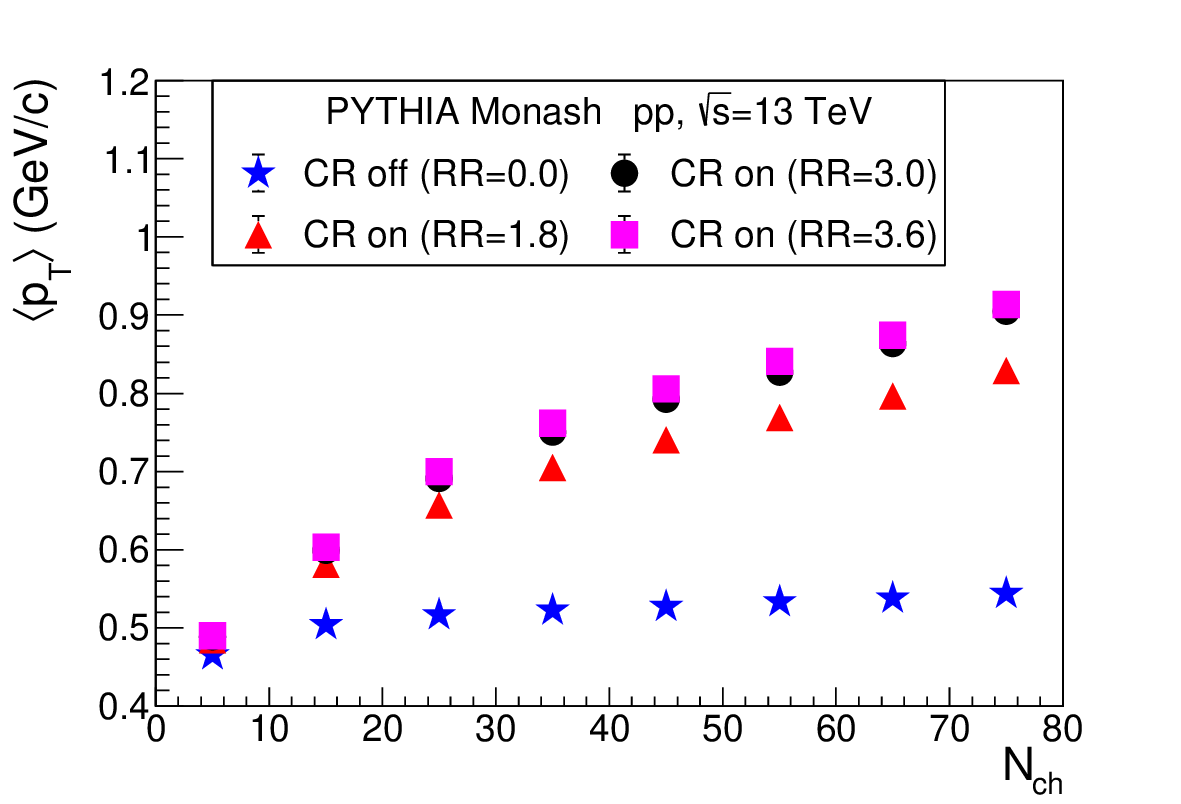}\label{fig:mpt2}}
	\caption{Variation of (a) effective temperature and (b) average $p_{\mathrm{T}}$ as a function of charged-particle multiplicity in pp collisions at $\sqrt{s}=$ 13 TeV for different values of reconnection range (RR) parameter in PYTHIA.}
	\end{figure}

	The variation of T$_{\rm {Eff}}$ and $\langle p_{\mathrm{T}} \rangle$ with charged-particle multiplicity in pp collisions at $\sqrt{s}=$ 13 TeV for RR=0.0, 3.0 and 3.6 are shown in Figs. \ref{fig:temp2} $\&$ \ref{fig:mpt2}, respectively, and compared with the previously obtained results of PYTHIA with default RR (=1.8) value. From these figures, it is readily evident that both the T$_{\rm {Eff}}$ and $\langle p_{\mathrm{T}} \rangle$ almost remain constant starting from low to high multiplicity interval class, when, CR mechanism is switched off. This behavior suggests that the properties of the medium formed are independent of the charged-particle multiplicity produced in the collisions, when no CR mechanism is implemented in PYTHIA. On the contrary, with the increase of RR values (=3.0 and 3.6) in PYTHIA, the increasing behavior of T$_{\rm {Eff}}$ and $\langle p_{\mathrm{T}} \rangle$ is found to be slightly sharper than that of the default RR results. Furthermore, the plateau-like region in T$_{\rm {Eff}}$ vs. multiplicity plot for RR=3.0 and 3.6 is observed to occur at higher T$_{\rm {Eff}}$ than the default PYTHIA results. It is interesting to note that the increase of the T$_{\rm {Eff}}$ and $\langle p_{\mathrm{T}} \rangle$ with RR seems to slow down gradually for increasing RR parameters, and no significant difference could be observed between the results of RR=3.0 and 3.6. These behaviors indicate that the color reconnection mechanism in PYTHIA has a substantial role in the flattening of observables like T$_{\rm {Eff}}$ and $\langle p_{\mathrm{T}} \rangle$ for N$_{\rm ch}$$>$30. However, a further increase of CR's strength beyond a certain limit may not necessarily affect the behavior of the system considerably.

\begin{figure*}[htp]
	\centering
	\includegraphics[width=140mm, height=70mm]{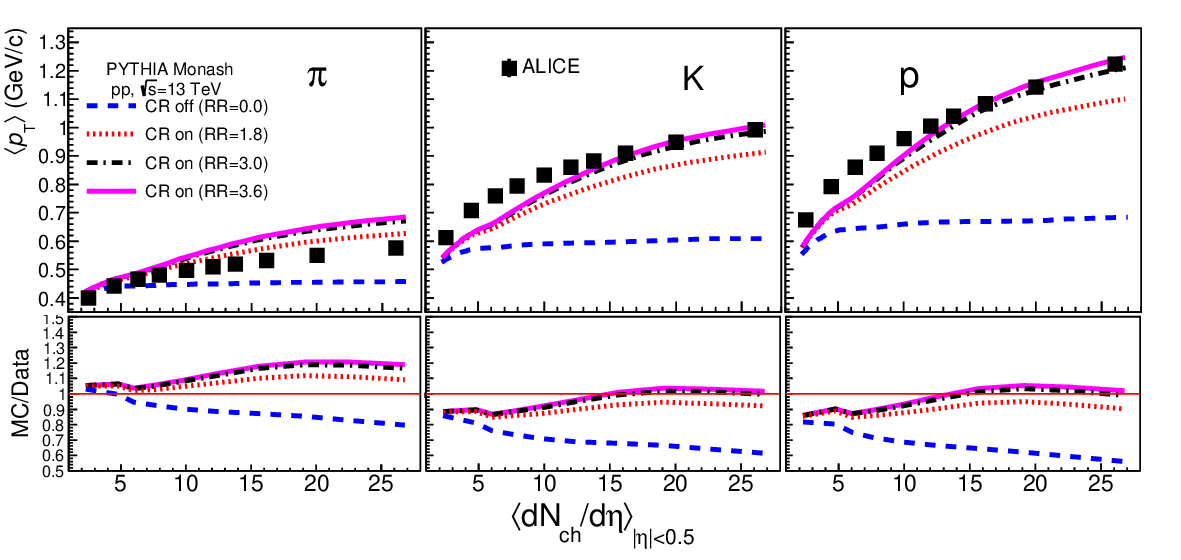}
	\caption{Pions, kaons and protons average $p_{\mathrm{T}}$ as a function of charged-particle multiplicity in pp collisions at $\sqrt{s}=$ 13 TeV for the PYTHIA Monash generated data with different values of reconnection range (RR) parameter are compared with the results of the ALICE data. }
	\label{fig:mpt_pikp}
	\end{figure*}

	Moreover, to gather some useful information about the collective-like behavior in small systems, further studies on pions, kaons, and protons average $p_{\mathrm{T}}$ ($\langle p_{\mathrm{T}} \rangle$) have been carried out in pp collisions at $\sqrt{s}=$ 13 TeV with the same sets of PYTHIA Monash generated data. It has recently been observed that pions, kaons, and protons $\langle p_{\mathrm{T}} \rangle$ increase with the increasing multiplicity, and the effect is more pronounced for heavier particle, i.e., for protons \cite{acharya_pikpvsmult}. This behavior is similar to that observed in the lower energy pp, p-Pb, and Pb-Pb collisions. In heavy-ion collisions, such a mass-dependent effect is considered to be a manifestation of the hydrodynamical evolution of the system \cite{abelev1}. However, for small colliding systems, the reasons for such mass-dependent enhancement of $\langle p_{\mathrm{T}} \rangle$ is beyond one's comprehension. To better understand the origin of such effect for pp collisions, studies with MC generated data were also performed and the results are reported in Ref. \cite{acharya_pikpvsmult}. It has been observed  in Ref. \cite{acharya_pikpvsmult} that the PYTHIA model with color reconnection mechanism (default) can describe the ALICE results on $\langle p_{\mathrm{T}} \rangle$ for pions quantitatively whereas, only a qualitative evolution of kaons and protons $\langle p_{\mathrm{T}} \rangle$ with increasing multiplicity could be observed. For a more quantitative description of experimental $\langle p_{\mathrm{T}} \rangle$ evolution with multiplicity and particle mass, as well as to have better insight into the observed results of kaons $\&$ protons, in this work the analysis on $\langle p_{\mathrm{T}} \rangle$ has been carried out with the PYTHIA generated data of different CR strength. To make a better comparison with experimental results, the event classification in the PYTHIA generated data is performed on the number of charged particles produced in the pseudorapidity region 2.8$<$$\eta$$<$5.1 and $-$3.7$<$$\eta$$<$$-$1.7, in the same way as done in Ref. \cite{acharya_pikpvsmult}. Then, the mean charged-particle multiplicity is estimated $|\eta|$$<$0.5. The results estimated with different sets of data with RR=0.0, 1.8 (default), 3.0 and 3.6 are plotted in Fig. \ref{fig:mpt_pikp} and are compared with the published results of ALICE. It could be observed from the figure that for RR=3.0 and 3.6, pions $\langle p_{\mathrm{T}} \rangle$ are overestimated from the ALICE and default PYTHIA results. On the other hand, kaons and protons $\langle p_{\mathrm{T}} \rangle$ can be described well by the results of RR=3.0 and 3.6 for charged-particle multiplicity density greater than 13. However, for multiplicity density less than 13, although it underestimates ALICE results, the agreement is still better than the default RR value. These results suggest that the color reconnection mechanism can surely mimic the mass-dependent hardening of $\langle p_{\mathrm{T}} \rangle$, similar to those observed in experimental pp collisions. This flow-like effect, produced by the CR does not require thermalisation or the formation of a medium. Moreover, the PYTHIA model results in different CR's strengths could not provide a quantitative and simultaneous description of $\langle p_{\mathrm{T}} \rangle$ of $\pi$, K and p in all multiplicities. This behavior indicates that the boost generated by the CR mechanism may not be the only cause of mass-dependent hardening of $\langle p_{\mathrm{T}} \rangle$ that is observed in experimental data. Also, these results provide a hint that mass-dependent hardening of $\langle p_{\mathrm{T}} \rangle$ for $\pi$, K and p with increasing multiplicity can't be considered a clear signal of the hydrodynamic evolution of the system.	

\section{Summary}
Transverse momentum spectra of the produced primary charged particles have been studied in pp collisions at the LHC energies using the Monte Carlo event generator PYTHIA Monash and are compared with the pseudorapidity and $p_{\rm T}$ spectra of the ALICE Collaboration for $\sqrt{s}=$ 7 and 13 TeV. The results of default PYTHIA Monash generated data are found to be in good agreement with the experimental data. For PYTHIA generated data, the $p_{\mathrm{T}}$ spectra are observed to flatten with increasing charged-particle multiplicity. The variations of effective temperature and the average transverse momentum of the system as a function of multiplicity exhibit a plateau-like region in T$_{\rm {Eff}}$ and $\langle p_{\mathrm{T}} \rangle$ with increasing multiplicity. Since PYTHIA does not include QCD type phase-transition, the appearance of a plateau-like region with PYTHIA data hints that the aforesaid de-confinement transition may not necessarily be the only mechanism that could give rise to such an effect. A further study with PYTHIA-generated data shows that the T$_{\rm {Eff}}$ and $\langle p_{\mathrm{T}} \rangle$ have no significant variation with charged-particle multiplicity when CR is switched off. On the other hand, with the increase of color reconnection strength, i.e., reconnection range (RR) value, the variations of T$_{\rm {Eff}}$, and $\langle p_{\mathrm{T}} \rangle$ against charged-particle multiplicity increase initially more sharply and then gradually saturates. No significant CR strength dependence variation of T$_{\rm {Eff}}$ and $\langle p_{\mathrm{T}} \rangle$ could be observed for RR=3.0 and 3.6. Thus, from this study, it is evident that the color reconnection mechanism in PYTHIA has a significant effect on inclusive charged particles $p_{\mathrm{T}}$ spectra, T$_{\rm {Eff}}$, and $\langle p_{\mathrm{T}} \rangle$ up to a certain threshold value of RR. The observed collective-like behavior such as flattening of T$_{\rm {Eff}}$ and $\langle p_{\mathrm{T}} \rangle$ for N$_{\rm ch}$$>$30 could be a manifestation of the color reconnection between the partons produced in hard multi-particle interaction. Further studies on identified particles such as pions, kaons, and protons' $\langle p_{\mathrm{T}} \rangle$ with different values RR parameter of PYTHIA suggest that even though the color reconnection mechanism can mimic the mass-dependent hardening of $\langle p_{\mathrm{T}} \rangle$ as observed in the ALICE experimental pp data, a simultaneous and quantitative description of the same is not possible with different strengths of the CR mechanism. These differences between the results of the MC and experimental data suggest that although the CR mechanism can mimic several collective-like effects as observed in the heavy-ion collisions, it may not be the only cause of collective-like behaviour observed in experimental data of pp collisions.

\section*{Acknowledgments}

The authors thankfully acknowledge the efforts of the PYTHIA Monash MC event generator team and making it freely available. The authors also acknowledge the Department of Science and Technology (DST), Government of India, for providing funds via Project No. SR/MF/PS-01/2014-GU(C) to develop a high-performance computing cluster (HPCC) facility for the generation of the Monte Carlo events for this work. P.S. would also like to thank DST, Government of India, for providing a scholarship in the form of JRF.



\end{document}